\documentclass[twocolumn,nofootinbib,amsmath,amssymb,a4paper]{revtex4}

\usepackage{graphicx}
\usepackage{dcolumn}
\usepackage{bm}

\input babarsym

\def\Dztilde   {\ensuremath {\tilde{D}^0}\xspace}
\def\bea{\begin{eqnarray}}
\def\eea{\end{eqnarray}}
\def\dodstartilde {\ensuremath {\tilde{D}^{(\ast)0}}\xspace}
\def\K  {\ensuremath{K}\xspace}

\newcommand{\npa}       [1]  {\npBase\ A~{\bf #1}}

\def\bea{\begin{eqnarray}}
\def\eea{\end{eqnarray}}

\newcommand{\re}{\ensuremath{\mathop{\rm Re}}}
\newcommand{\im}{\ensuremath{\mathop{\rm Im}}}

\newcommand{\fis}{\ensuremath{\mbox{$\mathcal{F}$}}\xspace}

\def \xsp {\ensuremath {x_{s+}}\xspace}
\def \xsm {\ensuremath {x_{s-}}\xspace}

\def \ysp {\ensuremath {y_{s+}}\xspace}
\def \ysm {\ensuremath {y_{s-}}\xspace}

\begin{document}

\title{Measurement of the CKM angle $\gamma$ with $B^\mp\rightarrow D^{(*)}[K^0_s\pi^-\pi^+]K^{(*)\mp}$ decays in \babar
\footnote{Proceedings of the 4th International Workshop on the CKM Unitarity Triangle (CKM'06), December 12-16, 2006, Nagoya (Japan).}
}

\author{F. Mart\'{\i}nez-Vidal}
\email{martinef@slac.stanford.edu}
\affiliation{%
IFIC, Universitat de Val\`encia-CSIC, E-46071 Valencia, Spain\\
{\rm (On behalf of the BaBar Collaboration)}
\\
}%

\begin{abstract}
We report on the measurement of the Cabibbo-Kobayashi-Maskawa angle $\gamma$ through
a Dalitz analysis of neutral $D$ decays to $\KS \pim \pip$ in the processes
$B^\mp\rightarrow D^{(*)} \Kmp$ and $B^\mp\rightarrow D K^{*\mp}$, $D^* \to D \piz, D \gamma$,
with the \babar\ detector at the SLAC PEP-II $e^+ e^-$ asymmetric-energy collider.
\end{abstract}

\maketitle

\section{Introduction and overview}
The angle $\gamma$ of the unitarity triangle is the phase of the Cabibbo-Kobayashi-Maskawa (CKM)
matrix~\cite{ref:CKM} defined as $\gamma\equiv\arg{\left[-V_{ud}^{}V_{ub}^{*}/V_{cd}^{}V_{cb}^{*}\,\right]}$,
which corresponds to the phase of the element $V^*_{ub}$, i.e. $V_{ub} = |V_{ub}|e^{-i\gamma}$,
in the Wolfenstein parameterization~\cite{ref:wolfenstein}. The precise measurement of the angle $\gamma$ is
a crucial goal of the physics program at the B-factories, however, it is also one of the most
difficult to achieve.

Among all methods proposed to extract $\gamma$, only those
using $B^\mp\to \Dztilde K^\mp$ decays are theoretically clean because the main contributions to the
amplitudes come from tree-level diagrams
(the symbol \Dztilde indicates either a $D^0$ or a $\bar{D}^0$ meson). 
The interference between the color allowed 
$B^-\to D^0K^-$ ($\b \to \c\ubar\s$) and the color suppressed 
$B^-\to \bar{D^0}K^-$ ($\b \to \u \cbar \s$) transitions~\cite{ref:chargeconj},
when the $D^0$ and $\bar{D^0}$ are reconstructed
in a common final state~\cite{ref:gronau,ref:soni,ref:ggsz_ads,ref:belle_dal04}, introduces a relative phase $\gamma$
in the decay amplitude. 
The sensitivity to $\gamma$ depends on the magnitude of the ratio of the $\b \to \u \cbar s$ amplitude with respect to
the $\b \to \c\ubar\s$ one, $r_B$, which plays a key role on the ability to measure $\gamma$ at the B-factories.
Theoretical expectations, consistent with current experimental limits, give 
$r_B \approx \mid V_{ud}^{}V_{ub}^{*}/V_{cd}^{}V_{cb}^{*} \mid c_F \sim 0.1$, where $c_F\sim0.2$ is the color 
suppression factor.


When the \Dztilde is reconstructed in a 3-body final state 
like $\KS \pim \pip$, the
interference between doubly-Cabibbo suppressed, Cabibbo allowed and \CP-eigenstate amplitudes provides 
strong phases to ensure the sensitivity to $\gamma$~\cite{ref:ggsz_ads,ref:belle_dal04}. The angle $\gamma$ can then be extracted through an
analysis of the distribution of the events in the \Dztilde Dalitz plane.

Assuming negligible effects from $\Dz-\Dzb$ mixing~\cite{ref:dmixing} and \CP asymmetries~\cite{ref:dcpv} in $D$ decays,
the $\Bmp \to \tilde{D}^{(*)0} \Kmp$, 
with $\tilde{D}^{*0} \rightarrow \Dztilde\pi^0,\Dztilde\gamma$,
$\Dztilde \to \KS \pim \pip$ decay chain rate 
$\Gamma^{(*)}_{\mp}(m^2_-,m^2_+)$ 
can be written as
\bea
\Gamma_{\mp}^{(*)}(m^2_-,m^2_+) \propto |{\cal A}_{D\mp}|^2+r_{B}^{(*)^2}|{\cal A}_{D\pm}|^2 + \ \ \ \ \ \ \ \ \ \ \ \ \ \ \ \ \nonumber  \\ 
   2 \epsilon \left\{ x_{\mp}^{(*)} \re[{\cal A}_{D\mp} {\cal A}^*_{D_\pm}] + 
             y_{\mp}^{(*)} \im[{\cal A}_{D\mp} {\cal A}^*_{D\pm}] \right\},~\label{eq:ampgen}
\eea
where $m^2_-$ and $m^2_+$ are the squared invariant masses of the $\KS\pim$ and $\KS\pip$ combinations, respectively,
${\cal A}_{D\mp} \equiv {\cal A}_{D}(m^2_\mp,m^2_\pm)$,
with ${\cal A}_{D-}$ (${\cal A}_{D+}$) the amplitude of the $\Dz \to \KS\pim\pip$ ($\Dzb \to \KS\pip\pim$) decay.
We introduce the 
{\it \CP (cartesian)} parameters~\cite{ref:babar_dalitzpub} 
$x_{\mp}^{(*)} = r_{B}^{(*)} \cos(\delta_{B}^{(*)} \mp \gamma)$ 
and 
$y_{\mp}^{(*)} = r_{B}^{(*)} \sin(\delta_{B}^{(*)} \mp \gamma)$, 
verifying 
$x_{\mp}^{(*)^2} + y_{\mp}^{(*)^2} = r_{B}^{(*)^2}$.
Here, 
$r_{B}^{(*)}$ 
is the magnitude of the ratio of the amplitudes 
${\cal A}(\Bm \to \bar{D}^{(*)0} \Km)$ 
and
${\cal A}(\Bm \to D^{(*)0} \Km)$ 
and 
$\delta^{(*)}_{B}$ 
is their relative strong phase. 
The factor $\epsilon$ in Eq.~(\ref{eq:ampgen}) takes the value $-1$
for the decay $\Bmp \to \tilde{D}^{*0}[\Dztilde\gamma] \Kmp$ and $+1$ for all the rest.
This relative sign arises due to parity and angular momentum conservation in the \dodstartilde decay,
and the different \CP content of $\Dztilde\gamma$ with respect to $\Dztilde\piz$~\cite{ref:bondar_gershon}.

Equation~(\ref{eq:ampgen}) also applies to $\Bmp \to \Dztilde \Kstarmp$ decays, with the replacements 
$r_{B}^{(*)} \to r_s$, $\delta_B^{(*)} \to \delta_s$,
$x_{\mp}^{(*)} \to x_{s\mp}= \kappa r_s \cos(\delta_s \mp \gamma)$, and
$y_{\mp}^{(*)} \to y_{s\mp}= \kappa r_s \sin(\delta_s \mp \gamma)$, verifying 
$x_{s\mp}^2 + y_{s\mp}^2 = \kappa^2 r_s^2$.
Here, the parameter $\kappa$ accounts for interference between resonant and non-resonant \Kstar decays, as
a consequence of the natural width of the \Kstar, with $0\leq \kappa\leq 1$~\cite{ref:gronau2002}. 
This general parameterization also accounts for variations of $r_s$ and $\delta_s$ within 
the \Kstar mass window, and for efficiency variations as a function of the kinematics of the \B decay.

\section{Data sample and event selection}

The analysis for $\Bm \to \tilde{D}^{(*)0} \Km$ ($\Bm \to \Dztilde \Kstarm$) decays~\cite{ref:chargeconj} is based on a sample of 
approximately 347 (227) million $B\bar B$ pairs collected by the \babar\ detector~\cite{ref:babar} at the SLAC PEP-II 
$e^+ e^-$ asymmetric-energy storage ring. For each signal $B$ decay channel we also reconstruct its own control sample,
$\Bm \to D^{(*)0} \pim$ ($\Bm \to \Dz a_1^-$).

The reconstruction and selection criteria are described in detail
elsewhere~\cite{ref:babar_dalitzpub,ref:babar_dalitzeps05,ref:babar_dalitzeps06}. 
$B$ meson candidates are characterized by using the energy difference $\Delta E$, the beam-energy substituted mass $\mes$,
and a Fisher discriminant \fis\ to separate $e^+e^-\to q\bar{q}$,  $q=u,d,s,c$ (continuum) and \BB events~\cite{ref:babar_dalitzpub}. 
If both $B^-\to \tilde{D}^{*0}[\Dztilde\pi^0]K^-$ and $B^-\to \tilde{D}^{*0}[\Dztilde\gamma]K^-$ candidates are selected in the 
same event, only the $B^-\to \tilde{D}^{*0}[\Dztilde\pi^0]K^-$ is kept. The cross-feed among the different samples is negligible 
except for $B^-\to \tilde{D}^{*0}[\Dztilde\gamma]K^-$, where the background from $B^-\to \tilde{D}^{*0}[\Dztilde\pi^0]K^-$ is 
about 5\% of the signal yield. This contamination has a negligible effect on the \CP parameters.

\begin{figure}[htb]
\begin{tabular}{cc}
\includegraphics[width=0.22\textwidth]{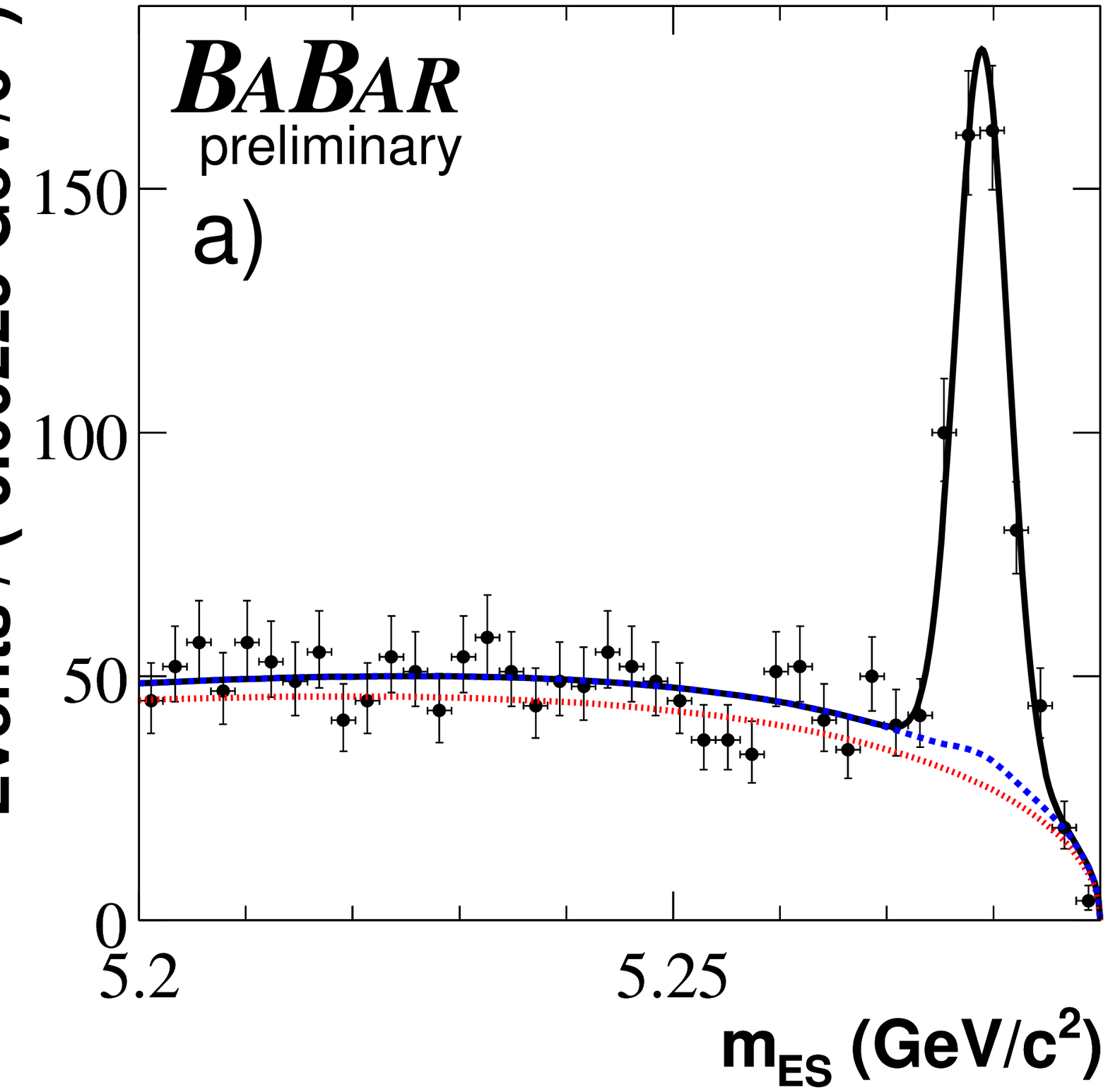}
\includegraphics[width=0.22\textwidth]{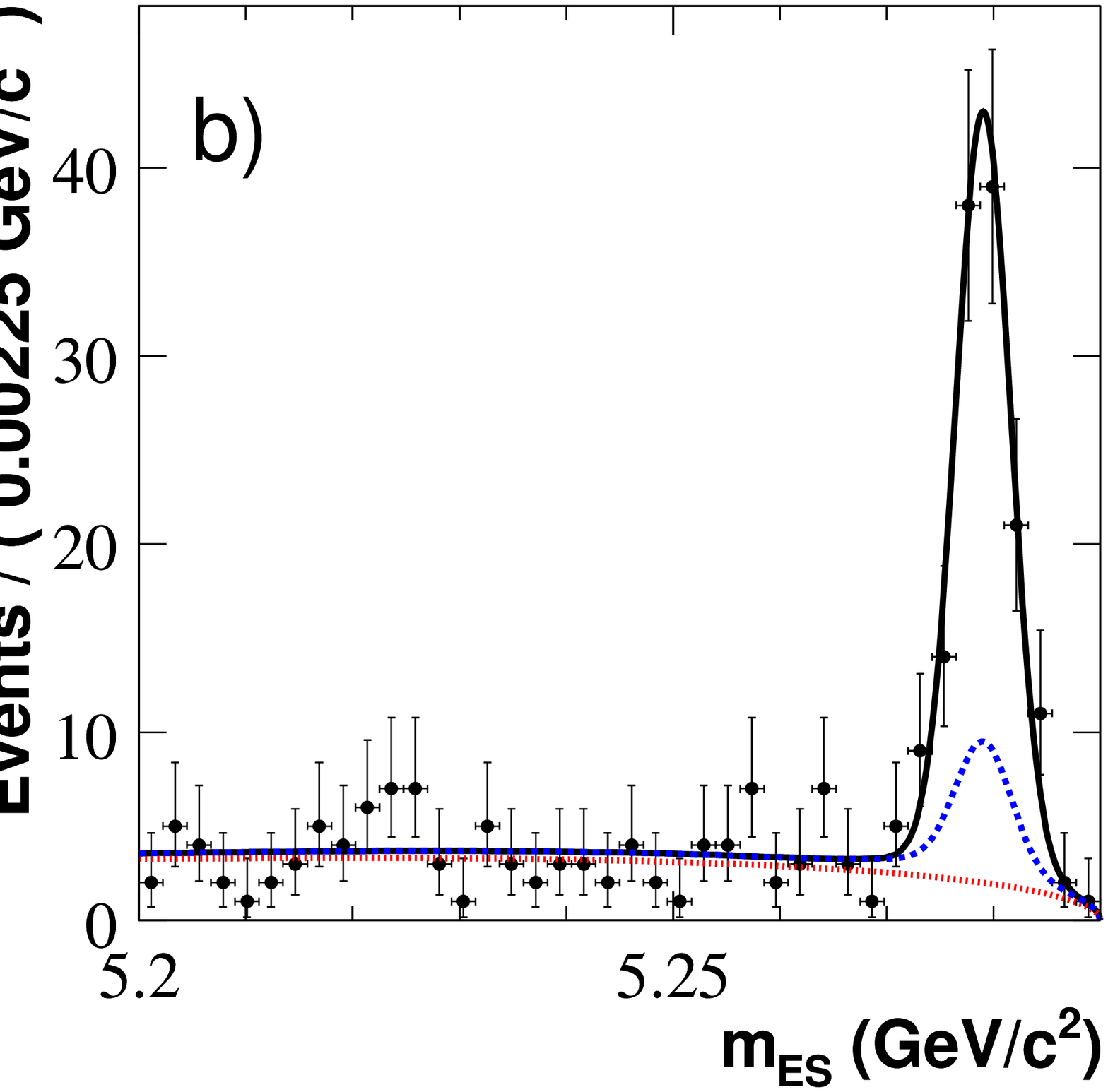} \\
\includegraphics[width=0.22\textwidth]{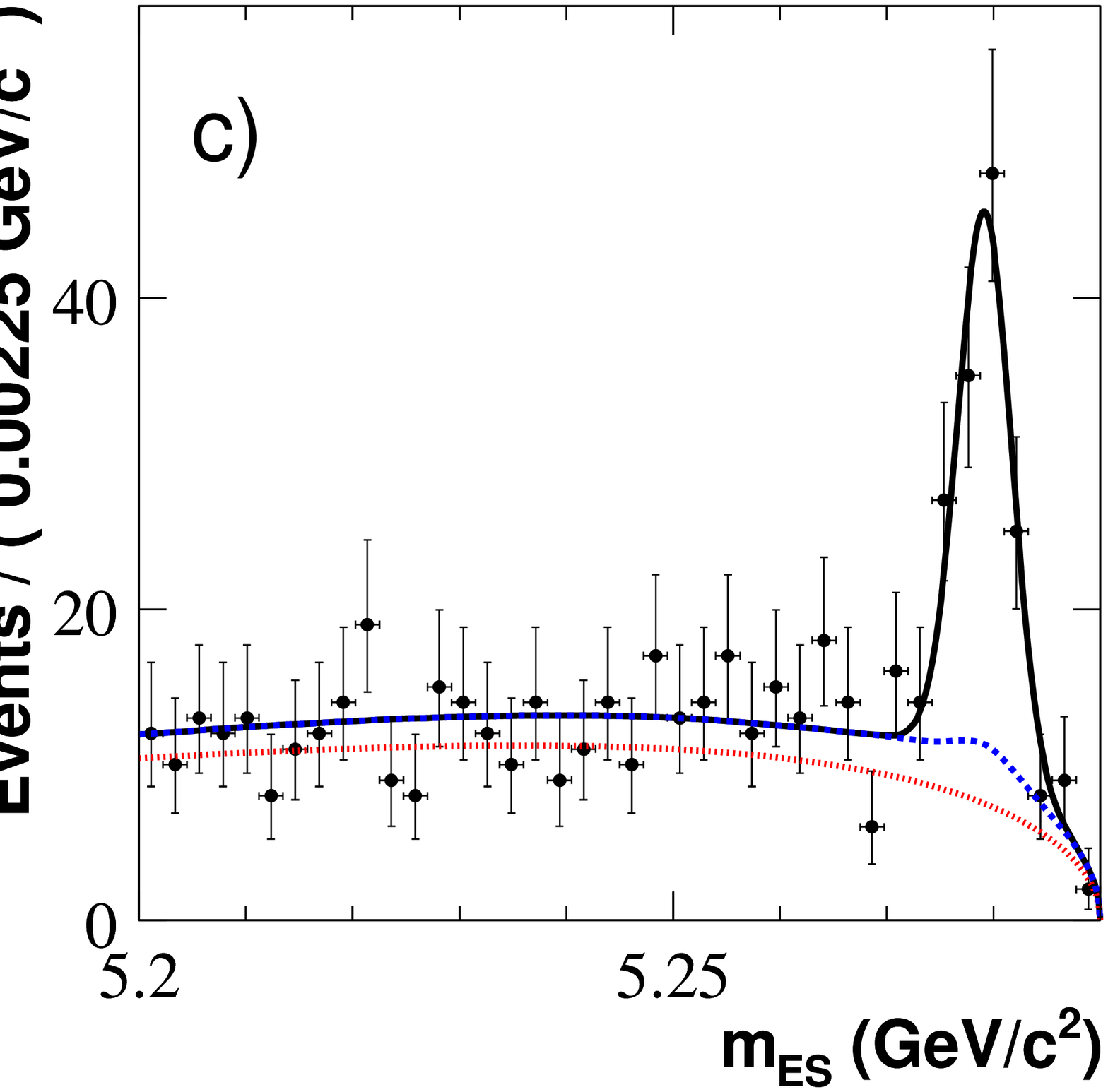}
\includegraphics[width=0.22\textwidth]{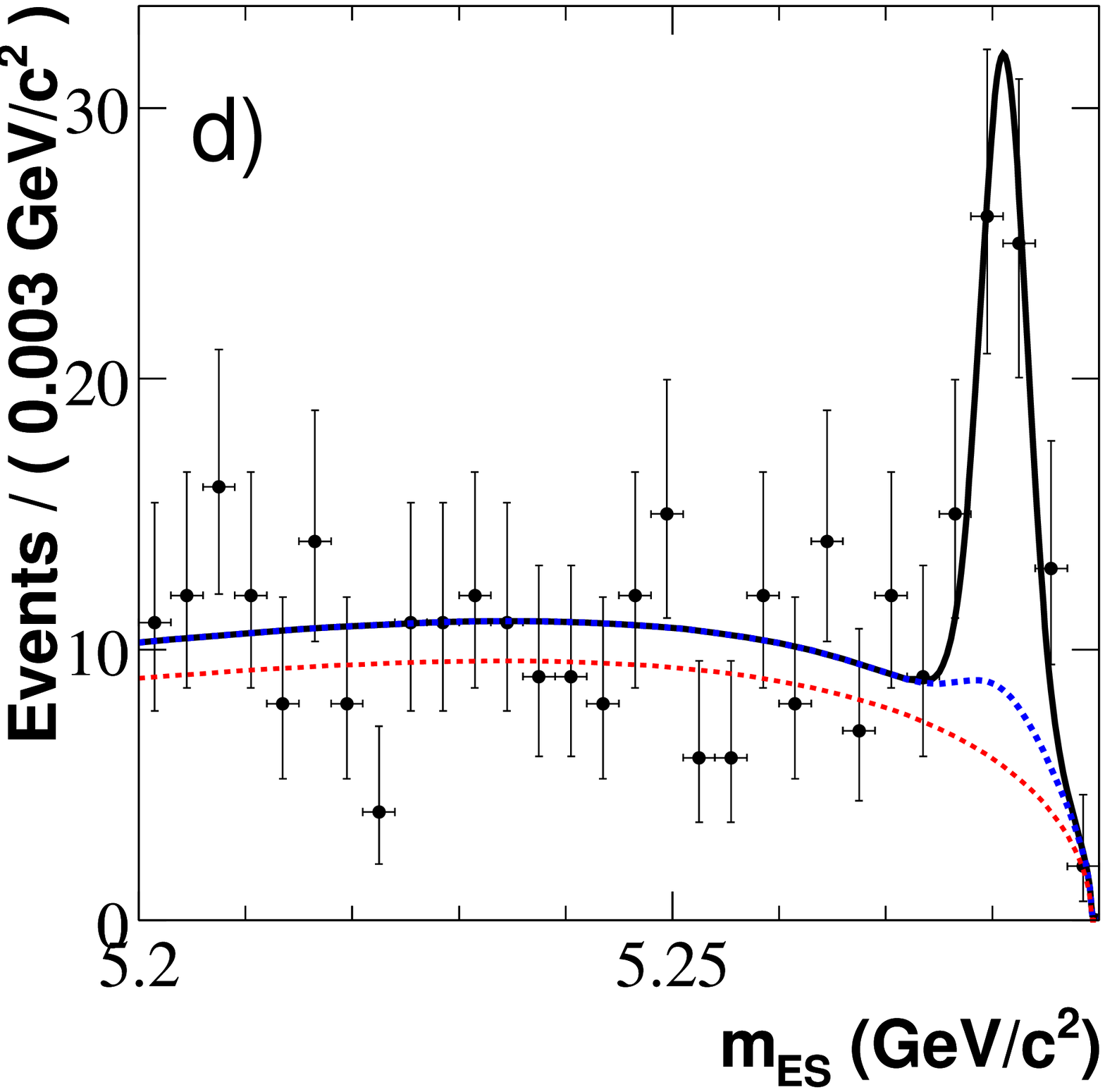}
\end{tabular}
\caption{\label{fig:btodk_mes} Distributions of \mes\ for (a) $B^-\to \Dztilde K^-$, 
(b) $B^-\to \tilde{D}^{*0}[\Dztilde\pi^0]K^-$, (c) $B^-\to \tilde{D}^{*0}[\Dztilde\gamma]K^-$,
and (c) $B^-\to \tilde{D}^{0}\Kstarm$.
The curves superimposed represent the overall fit projections (solid black lines), the continuum contribution (dotted red lines), 
and the sum of all background components (dashed blue lines).}
\end{figure}

The reconstruction efficiencies are $15\%$, $7\%$, $9\%$, and $11\%$, for the $B^-\to \Dztilde K^-$, 
$B^-\to \tilde{D}^{*0}[\Dztilde\pi^0]K^-$, $B^-\to \tilde{D}^{*0}[\Dztilde\gamma]K^-$, and $B^-\to \tilde{D}^0 \Kstarm$ decay modes, respectively. 
Figure~\ref{fig:btodk_mes} shows the \mes\ distributions after all selection criteria,
for $|\DeltaE|<30(25)$~\mev, for $B^-\to \tilde{D}^{(*)0} K^-$($B^-\to \tilde{D}^{0} \Kstarm$).
The largest background contribution comes from continuum events or \BB decays where a fake or true $D^0$ is combined with a random track.
Another source of background for $B^-\to \tilde{D}^{(*)0} K^-$ is given by $B^-\to D^{(*)0}\pi^-$ decays where the prompt 
pion is misidentified as kaon. These decays are separated from the signal using their different \DeltaE distribution.

\section{The $D^0\rightarrow \KS \pi^-\pi^+$ decay model}
\label{sec:dalitzModel}

\begin{figure}[htb]
\begin{tabular} {cc}  
{\includegraphics[width=0.2\textwidth]{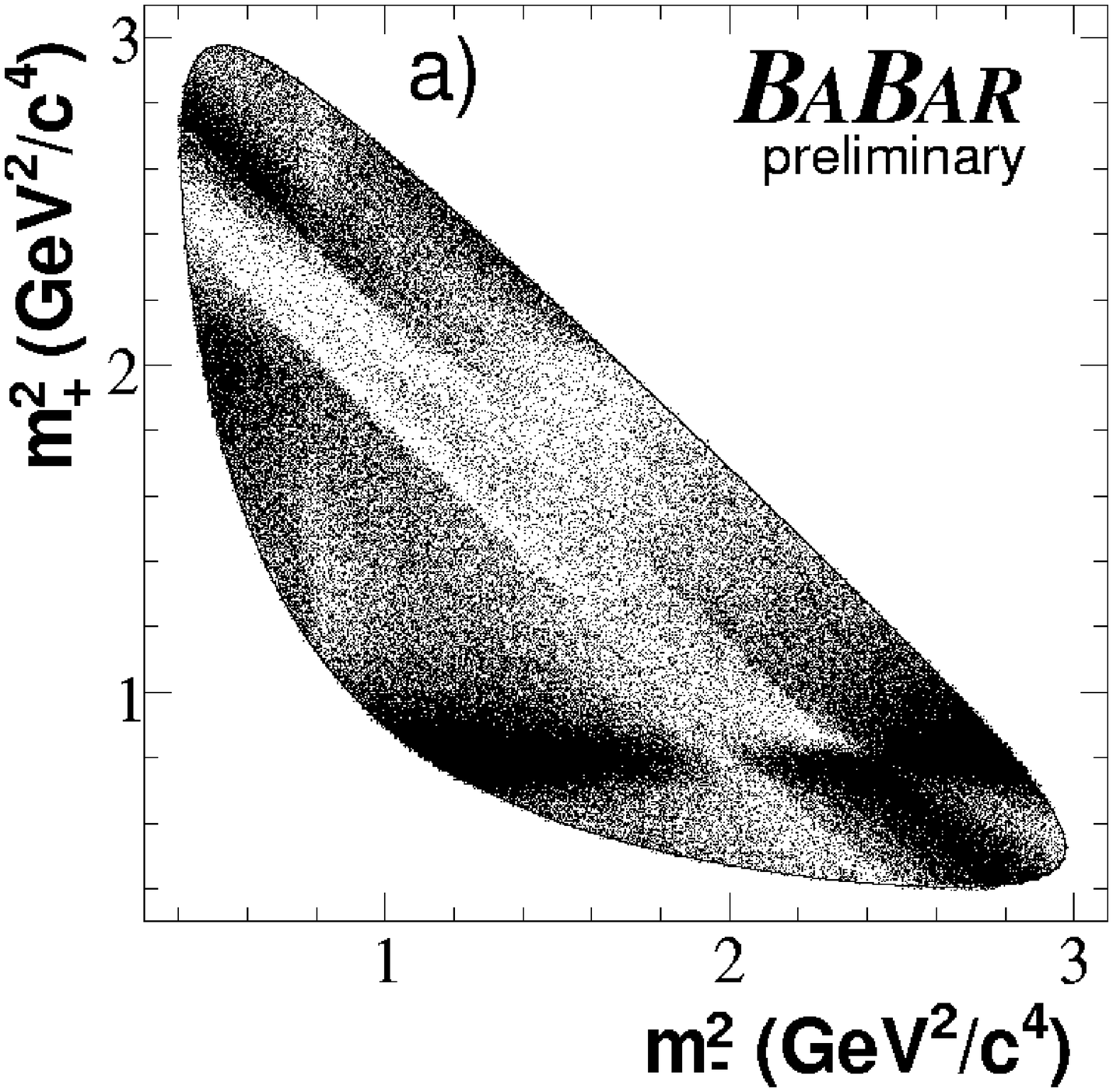}} &
{\includegraphics[width=0.2\textwidth]{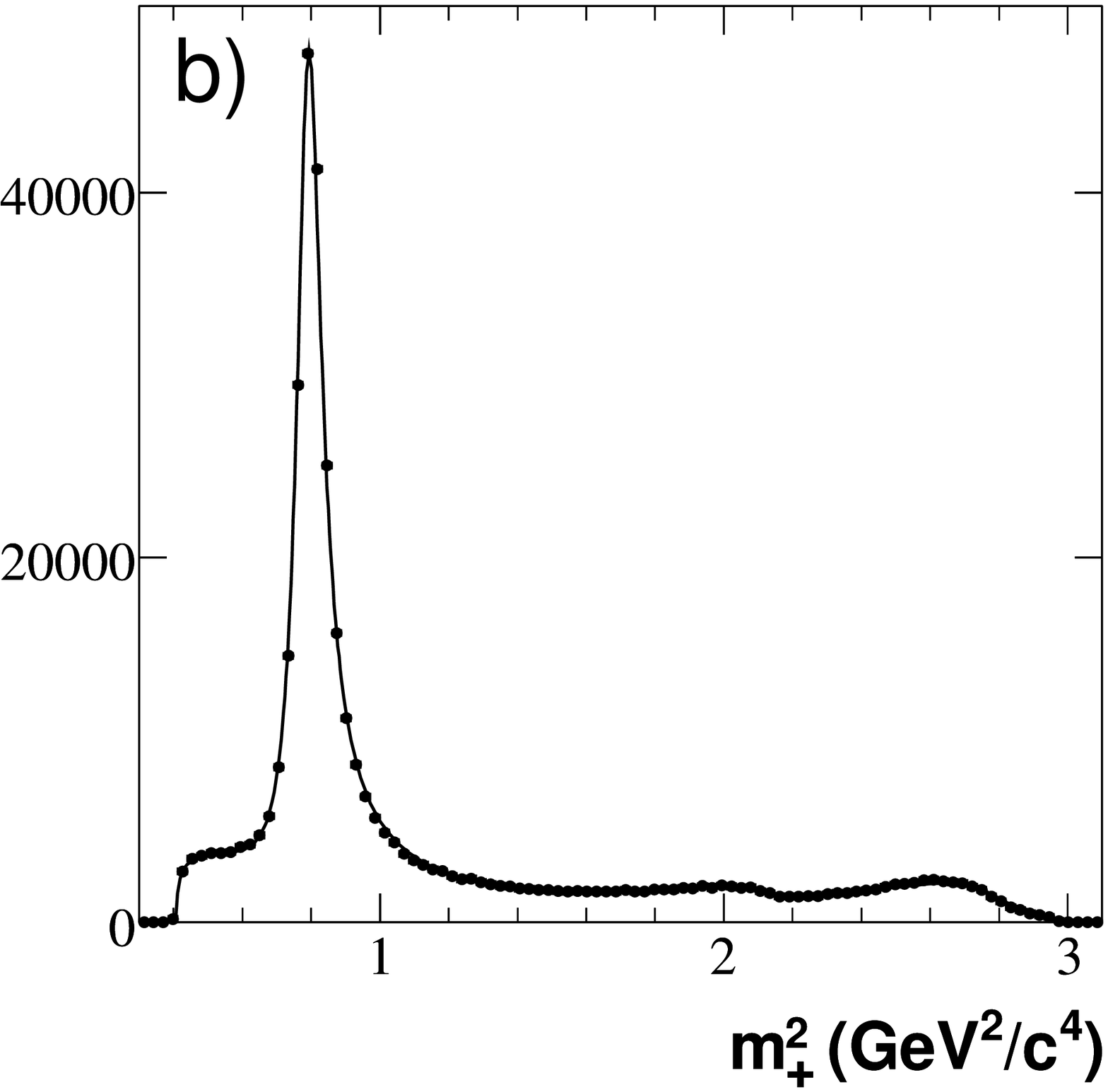}} \\
{\includegraphics[width=0.2\textwidth]{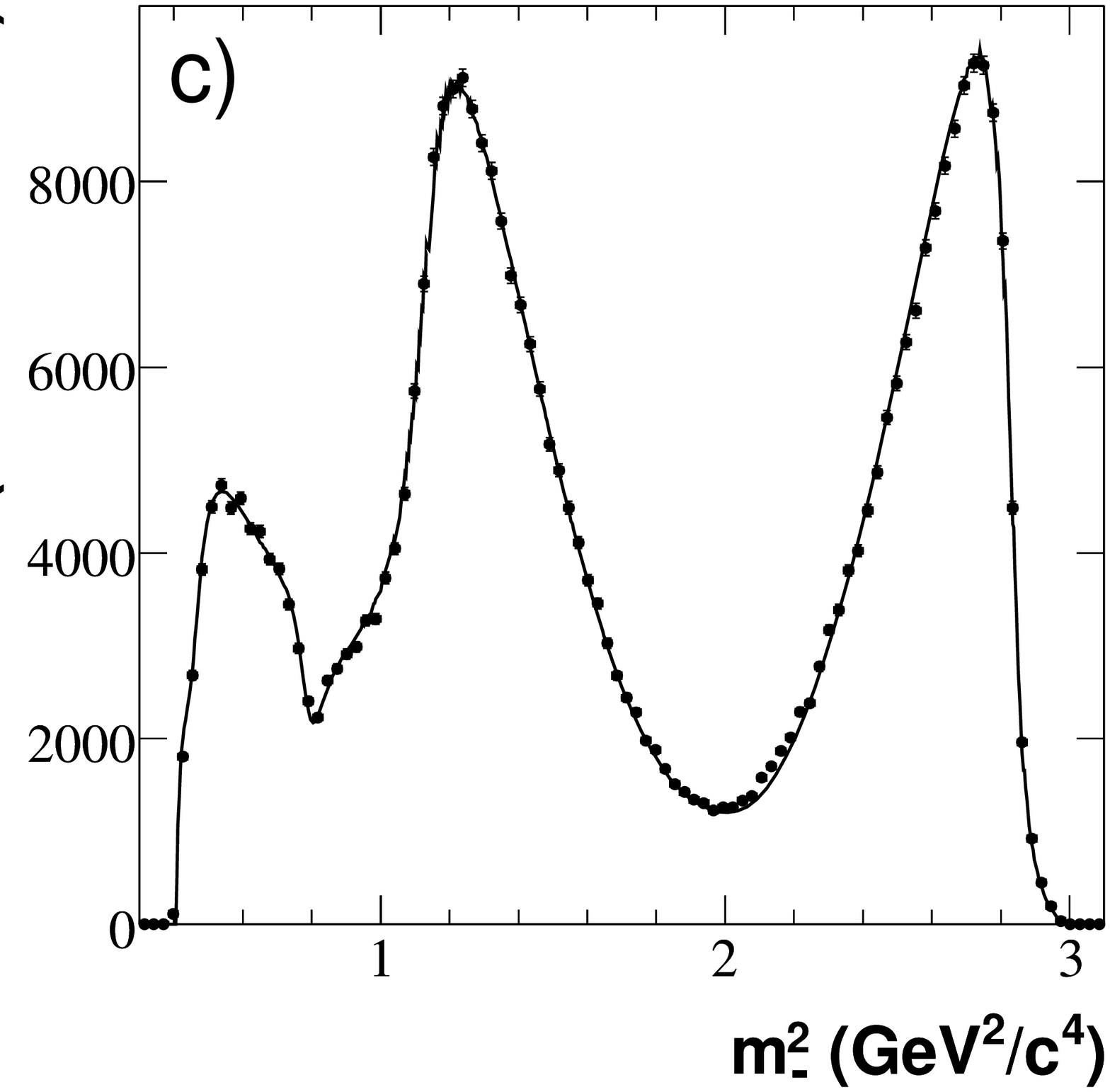}} &
{\includegraphics[width=0.2\textwidth]{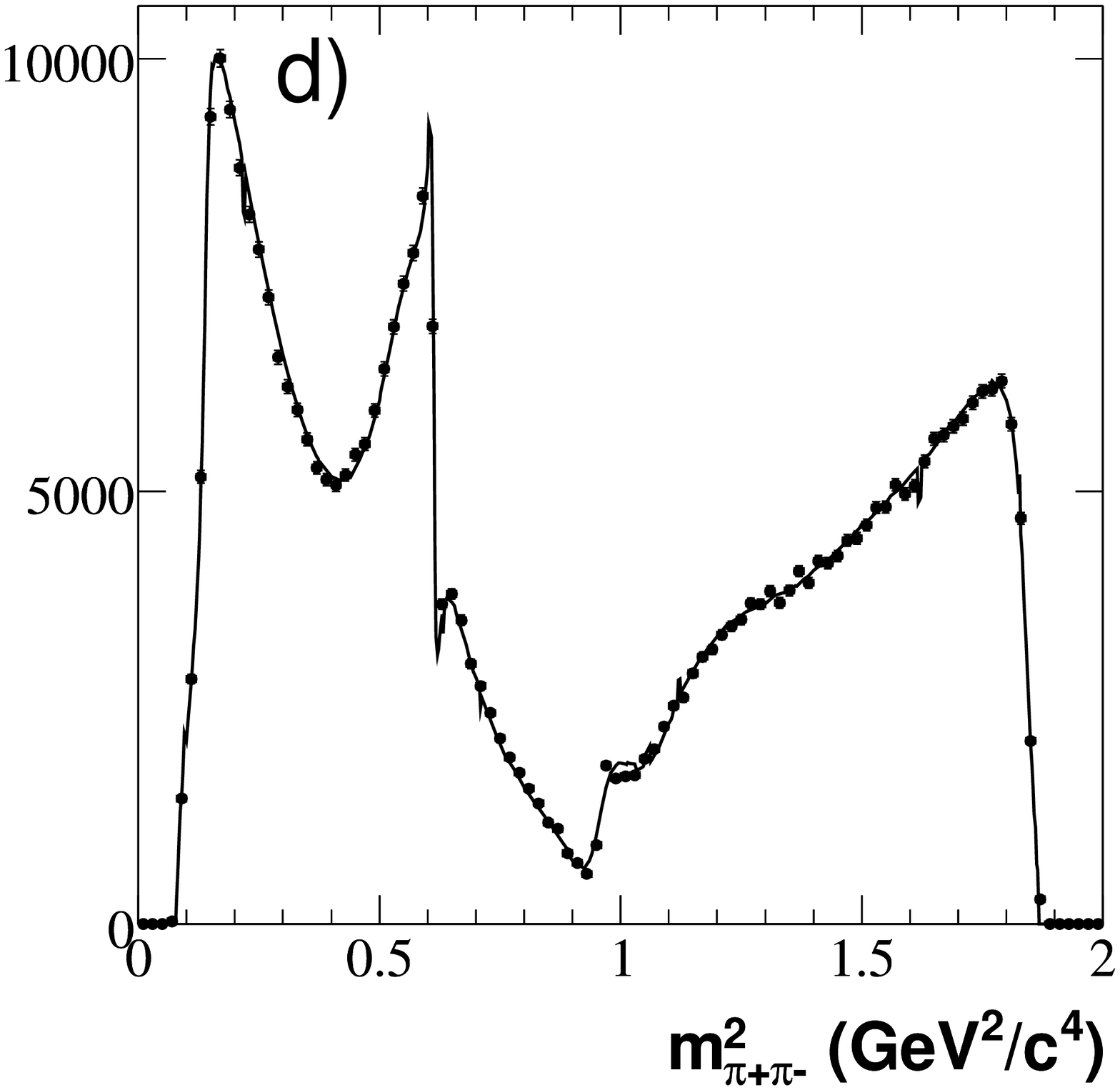}} \\
\end{tabular}   
\caption{\label{fig:BWfit-res} (a) The $\bar{D}^0 \to \KS \pim \pip$ Dalitz distribution from $D^{*-} \to \bar{D}^0 \pi^-$ events, and projections
on (b) $m^2_+=m^2_{\KS\pi^+}$, (c) $m^2_-=m^2_{\KS\pi^-}$, and (d) $m^2_{\pip\pim}$. $\Dz \to \KS \pi^+ \pi^-$ from $\Dstarp \to \Dz \pip$ events 
are also included. The curves are the reference model fit projections. }
\end{figure}

The $\Dz \to \KS \pim \pip$ decay amplitude ${\cal A}_D(m^2_-,m^2_+)$ is determined from an unbinned maximum-likelihood fit to
the Dalitz plot distribution of a high-purity (97.7\%) tagged \Dz sample from 390328 $\Dstarp\to\Dz\pip$ decays reconstructed in 270 \invfb of data, 
shown in Fig.~\ref{fig:BWfit-res}.  Our phenomenological reference model to describe ${\cal A}_D(m^2_-,m^2_+)$ uses a sum 
of two-body amplitudes (subscript $r$) and a non-resonant (subscript NR) contribution,
\bea
{\cal A}_D(m^2_-,m^2_+) = \sum_r a_r e^{i \phi_r} {\cal A}_r(m^2_-,m^2_+) + a_{\rm NR} e^{i \phi_{\rm NR}},
\eea
where the parameters $a_r$ ($a_{\rm NR}$) and $\phi_r$ ($\phi_{\rm NR}$) are the magnitude and phase of the amplitude for 
component $r$ (NR). The function ${\cal A}_r = F_r \times T_r \times W_r$ is the Lorentz-invariant expression that describes the
dynamic properties of the \Dz meson 
decaying into $\KS \pim \pip$ through an intermediate resonance $r$, as a function of position in the Dalitz plane. 
Here, $F_r$ is the Blatt-Weisskopf centrifugal barrier factor for the resonance 
decay vertex~\cite{ref:blatt-weisskopf} with radius $R=1.5$~GeV$^{-1}$~(0.3~\fm), 
$T_r$ is the resonance propagator, and $W_r$ describes the angular distribution in the decay.
For $T_r$ we use a relativistic Breit-Wigner (BW) parameterization, except for $r=\rho(770)$ and $\rho(1450)$ where
we use the functional form suggested in Ref.~\cite{ref:gounarissakurai}.
The angular dependence $W_r$ is described with the helicity formalism as 
shown in \cite{ref:cleo}\footnote{The label A and B should be swapped in Eq.~(6) of \cite{ref:cleo}.}. 
Mass and width values are taken from~\cite{ref:pdg2004}, with the exception of $K^{*}_0(1430)^+$ taken from \cite{ref:e791K*}.  
The model consists of a total of 13 resonances leading to 16 two-body decay amplitudes and phases 
(see Table~\ref{tab:BWfit-res}), and accounts
for efficiency variations across the Dalitz plane and the small background contribution.
All the resonances considered in this model are well established except for the two 
scalar $\pi\pi$  resonances, $\sigma$ and $\sigma'$, whose
masses and widths are obtained from our sample~\cite{ref:comment_sigma}. 
Their addition to the model is motivated by an improvement in the description of the data. 

The possible absence of the $\sigma$ and $\sigma'$ resonances is considered in the evaluation 
of the systematic errors through the use of a K-matrix formalism~\cite{ref:Kmatrix} to 
parameterize the $\pi\pi$ S-wave states. 
The K-matrix method provides a direct way of imposing
the unitarity constraint of the scattering matrix that is not guaranteed in the case of the BW model and is 
suited to the study of broad and overlapping resonances in multi-channel decays,
avoiding the need to introduce the two $\sigma$ scalars,
\bea
{\cal A}_D(m^2_-,m^2_+) = F_1(s) + \sum_{r\ne\pi\pi~S=0} a_r e^{i \phi_r} {\cal A}_r(m^2_-,m^2_+),~\label{eq:k-matrix}
\eea
where \mbox{$F_1(s)=\sum_j\left[I-iK(s)\rho(s)\right]^{-1}_{1j} P_j(s)$} is the contribution of $\pi\pi$ S-wave states.
Here, $s=m_{\pim\pip}^2$, $I$ is the identity matrix, $K$ is the matrix describing the S-wave
scattering process, $\rho$ is the phase-space matrix, and $P$ is the initial production
vector~\cite{ref:Kmatrix}. The index $j$ represents the $j^{\rm th}$ channel ($1=\pi\pi$, $2=\K\Kbar$, 
$3=$~multi-meson~\cite{ref:multimeson}, $4=\eta\eta$, $5=\eta\eta'$).
The K-matrix parameters are obtained from a global fit to the available
$\pi\pi$ scattering data below 1900~\mevcc~\cite{ref:AS}, while the initial production vector is obtained
from our fit to the tagged $\Dz \to \KS \pim \pip$ data.

\begin{table}
\caption{\label{tab:BWfit-res} Complex amplitudes $a_r e^{i\phi_r}$ and fit fractions of the different components ($K_S\pi^-$, 
$K_S\pi^+$, and $\pi^+\pi^-$ resonances) obtained from the fit of the $D^0 \to K_S\pi^-\pi^+$ Dalitz 
distribution from $D^{*+} \to D^0 \pi^+$ events. Errors are statistical only. 
The fit fraction is defined for the resonance terms as 
the integral of $a_r^2 |{\cal A}_r(m^2_-,m^2_+)|^2$ over the Dalitz plane divided by the integral
of $|{\cal A}_D(m^2_-,m^2_+)|^2$. The sum of fit fractions is
$119.5\%$. A value different from 100\% is a consequence of the interference among the amplitudes.}
\begin{ruledtabular}
\begin{tabular}{lccc}
\\[-0.15in]
    Component  &  $Re\{a_r e^{i\phi_r}\}$ &  $Im\{a_r e^{i\phi_r}\}$ & Fraction (\%) \\ [0.01in]
\hline \hline
$K^{*}(892)^-$        &  $-1.223\pm0.011$  &  $\phantom{-}1.346\pm0.010$  &  58.1 \\  
$K^{*}_0(1430)^-$     &    $-1.698\pm0.022$   &   $-0.576\pm0.024$  & 6.7 \\  
$K^{*}_2(1430)^-$     &    $-0.834\pm0.021$    &  $\phantom{-}0.931\pm0.022$  &  6.3 \\  
$K^{*}(1410)^-$       &   $-0.25\pm0.04$  &   $-0.11\pm0.03$    & 0.1 \\ 
$K^{*}(1680)^-$       &    $-1.285\pm0.014$     &   $\phantom{-}0.205\pm0.013$    &  0.6 \\ 
\hline
$K^{*}(892)^+$        &   $\phantom{-}0.100\pm0.004$   & $-0.127\pm0.003$   & 0.5 \\ 
$K^{*}_0(1430)^+$     &   $-0.027\pm0.016$    &    $-0.076\pm0.017$   & 0.0 \\ 
$K^{*}_2(1430)^+$     &   $\phantom{-}0.019\pm0.017$   &  $0\phantom{-}.177\pm0.018$   &  0.1 \\ 
\hline
$\rho(770)$           &     $1$     &      $0$                    &   21.6 \\ 
$\omega(782)$         &    $-0.0219\pm0.0010$  & $0.0394\pm0.0007$   & 0.7 \\ 
$f_2(1270) $          &    $-0.699\pm0.018$   &   $0.387\pm0.018$   & 2.1 \\ 
$\rho(1450)$          &    $\phantom{-0}0.25\pm0.04\phantom{0}$    &   $0.04\pm0.06$    &   0.1 \\ 
\hline 
Non-resonant          &    $-0.99\pm0.19\phantom{0}$   &     $\phantom{-}3.82\pm0.13$    &    8.5 \\ 
$f_0(980) $           &   $0.447\pm0.006$   &   $\phantom{-}0.257\pm0.008$   &  6.4 \\ 
$f_0(1370) $          &   $0.95\pm0.11$   &    $-1.619\pm0.011$      &  2.0 \\ 
$\sigma$              &    $1.28\pm0.02$   &  $\phantom{-}0.273\pm0.024$    &   7.6 \\ 
$\sigma '$            &  $0.290\pm0.010$   &   $-0.066\pm0.010$    &   0.9 \\ 
\end{tabular}
\end{ruledtabular}
\end{table}

\section{\CP fit results and systematic uncertainties}
\label{sec:cpfit}

Once the decay amplitude ${\cal A}_{D}(m^2_-,m^2_+)$ is known it can be fed into Eq.~(\ref{eq:ampgen}).
The extraction of the \CP-violating parameters $x_{\mp}^{(*)}$ and $y_{\mp}^{(*)}$ (\CP fit) is then performed
through a simultaneous maximum likelihood fit to the $\Gamma_{-}^{(*)}(m^2_-,m^2_+)$ and $\Gamma_{+}^{(*)}(m^2_-,m^2_+)$
Dalitz plot distributions for $\B^- \to \tilde{D}^{(*)0} \K^-$ and $\B^+ \to \tilde{D}^{(*)0} \K^+$ decays, respectively.
A similar fit is performed for $\B^\mp \to \tilde{D}^{0} \Kstarmp$ decays.
Different background components are considered: continuum, $B^-\to D^{(*)0}\pi^-$ for $B^-\to \tilde{D}^{(*)0} K^-$, and \BB.
The likelihood function uses the Dalitz plot distribution (after correction for efficiency variations), \mes, \DeltaE, and \fis,
with shapes determined directly from the signal and control samples, from both signal and sideband regions. 
Only the shapes for \BB background events are determined from Monte Carlo simulation.
Events falling into the continuum and \BB background components are
themselves divided into events with a real or a fake (combinatorial) \Dz. We finally account for the correlation
between the flavor of true \Dz mesons and the charge of combinational charged kaon.
 
\begin{figure}[htb]
\begin{tabular}{cc}
\includegraphics[width=0.22\textwidth]{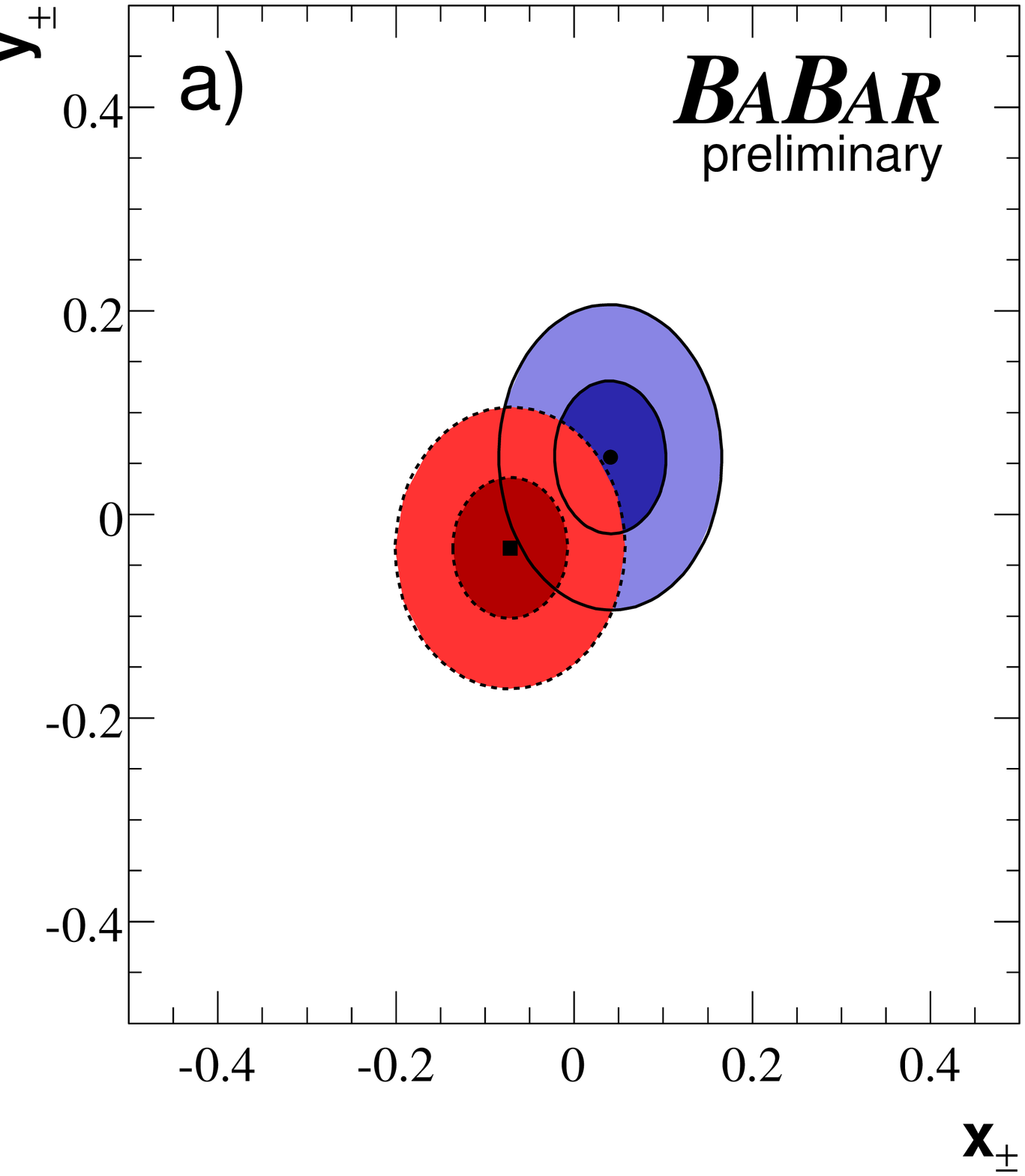}&
\includegraphics[width=0.22\textwidth]{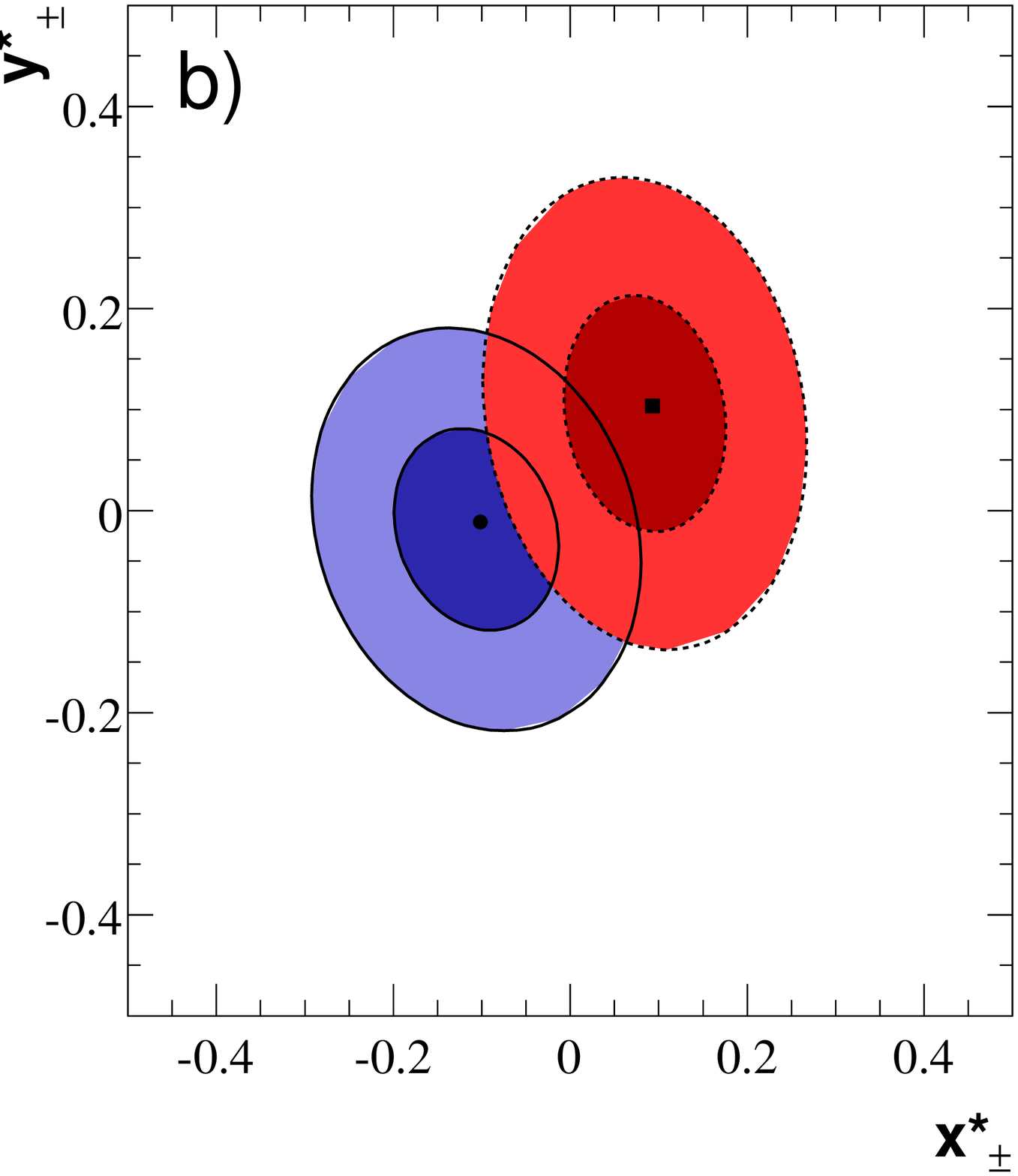}\\
\includegraphics[width=0.22\textwidth]{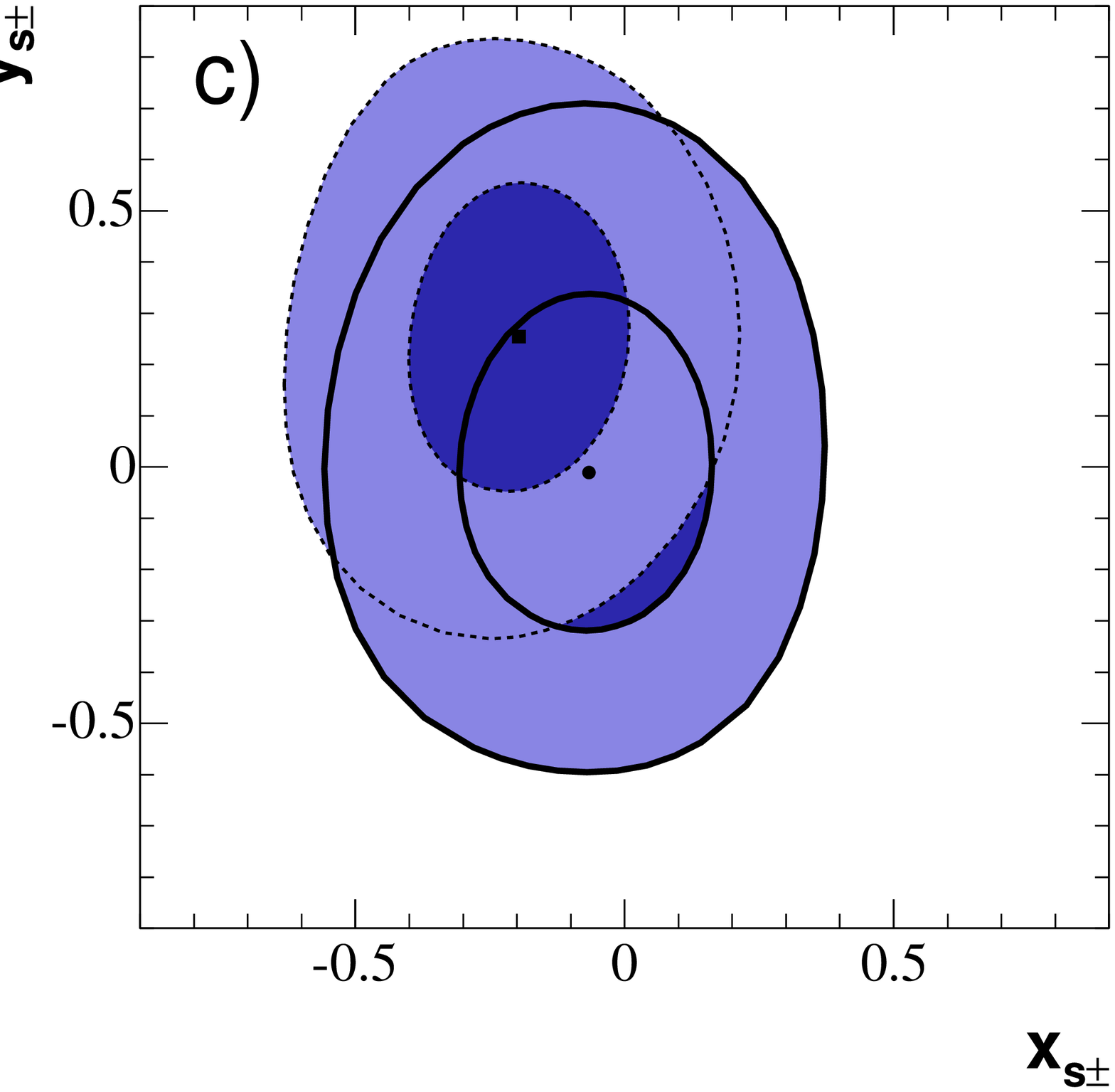}\\
\end{tabular}
\caption{\label{fig:cart_CL} 
Contours at $39.3\%$ (dark) and $86.5\%$ (light) confidence level (corresponding to two-dimensional one- and two-standard
deviation regions), 
including statistical and systematic uncertainties, for 
(a) $(x_{\mp},y_{\mp})$,
(b) $(x^{*}_{\mp},y^{*}_{\mp})$, and
(c) $(x_{s\mp},y_{s\mp})$ 
parameters, for \Bm (thick and solid lines) and \Bp (thin and dotted lines) decays.}
\end{figure}

We find $398\pm 23$, $97\pm 13$, $93\pm 12$, and $42\pm 8$ signal events, 
for $B^-\to \tilde{D}^0 K^-$, $B^-\to \tilde{D}^{*0}[\Dztilde\pi^0]K^-$, $B^-\to \tilde{D}^{*0}[\Dztilde\gamma]K^-$, and $B^-\to \tilde{D}^0 \Kstarm$,
respectively,
in agreement with expectations based on measured branching fractions and efficiencies estimated from Monte Carlo simulation. 
The results for the \CP-violating parameters $x^{(*)}_{\mp}$, $y^{(*)}_{\mp}$, $x_{s\mp}$, and $y_{s\mp}$,
are summarized in Table~\ref{tab:cp_coord}. 
The only non-zero statistical correlations involving the \CP parameters are 
for the pairs $(x_-,y_-)$, $(x_+,y_+)$, $(x_-^*,y_-^*)$, $(x_+^*,y_+^*)$, $(x_{s-},y_{s-})$, $(x_{s+},y_{s+})$,
which amount to $-1\%$, $1\%$, $-17\%$, $-14\%$, $-10\%$, and $2\%$, respectively.
Figure~\ref{fig:cart_CL} shows the one- and two-standard deviation confidence-level contours 
(including statistical and systematic uncertainties) in the $(x,y)$ plane for all the reconstructed modes, 
and separately for $B^-$ and $B^+$ decays. The separation of 
the $B^-$ and $B^+$ contours is an indication of direct \CP\ violation.

\begin{table*}
\caption{\label{tab:cp_coord} \CP-violating parameters $x^{(*)}_{\mp}$, $y^{(*)}_{\mp}$, $x_{s\mp}$, and $y_{s\mp}$, 
as obtained from the \CP\ fit. 
The first error is statistical, the second is experimental systematic uncertainty and the third is 
the systematic uncertainty associated with the Dalitz model.}
\begin{ruledtabular}
\begin{tabular}{lccc}
\\[-0.15in]
\CP\ parameter            & $\Bm \to \tilde{D}^0 K^-$ & $\Bm \to \tilde{D}^{*0} K^-$ & $\Bm \to \tilde{D}^0 \Kstarm$ \\ [0.01in] \hline
$x_{-}/x_{-}^{*}/x_{s-}$  & $\phantom{-}0.041\pm 0.059 \pm 0.018\pm 0.011$ & $-0.106\pm 0.091\pm 0.020\pm 0.009$ & $-0.20\pm0.20\pm0.11\pm0.03$ \\ 
$y_{-}/y_{-}^{*}/y_{s-}$  & $\phantom{-}0.056\pm 0.071\pm 0.007\pm 0.023$ & $-0.019\pm 0.096\pm 0.022\pm 0.016$ & $\phantom{-}0.26\pm0.30\pm0.16\pm0.03$ \\ 
$x_{+}/x_{+}^{*}/x_{s+}$  & $-0.072\pm 0.056 \pm 0.014\pm 0.029$ & $\phantom{-}0.084\pm 0.088\pm 0.015\pm 0.018$ & $-0.07\pm0.23\pm0.13\pm0.03$ \\
$y_{+}/y_{+}^{*}/y_{s+}$  & $-0.033\pm 0.066\pm 0.007\pm 0.018$ & $\phantom{-}0.096\pm 0.111\pm 0.032\pm 0.017$ & $-0.01\pm0.32\pm0.18\pm0.05$ \\ 
\end{tabular}
\end{ruledtabular}
\end{table*}

The largest single contribution to the systematic uncertainties on the \CP\ parameters comes from the choice of the Dalitz model 
used to describe the $D^0\to\KS\pi^-\pi^+$ decay amplitude. We use a set of alternative models where some resonances are removed 
or the parameterization of the different amplitudes are changed. For the $\pi\pi$ S-wave we use the K-matrix
approach described in Sec.~\ref{sec:dalitzModel}, while for the P-wave we change the mass and width 
describing the $\rho(770)$ within their quoted uncertainty~\cite{ref:pdg2004}. 
The uncertainty on the description of the $K\pi$ S-wave is estimated by floating in our flavor tagged \Dz sample
the mass and width of the BW describing the $K^*(1430)$, and using an additional parameterization taken from Ref.~\cite{ref:lassparam1} 
with parameters extracted from our fit. Since the $K\pi$ P-wave is dominated by the $K^*(892)$ in both
Cabibbo allowed and doubly Cabibbo suppressed amplitude, the mass and the width of this resonance, taken from \cite{ref:pdg2004} in 
the reference model, are changed to the values obtained from our fit to the tagged \Dz sample. The
resulting values are consistent with what is found in $B\to J/\Psi K\pi$ decays selected in \babar\ data.
For the $\pi\pi$ and $K\pi$ D-waves, described by the $f_2(1270)$ and $K^*_2(1430)$ resonances, respectively,
we use as alternative 
the formalism derived 
from Zemach tensors~\cite{ref:zemach}. The difference is very small for P-waves but is larger for D-waves.  
Other alternative models are built by removing the Blatt-Weisskopf penetration factors~\cite{ref:blatt-weisskopf},
removing resonances with small fit fractions --$K^*_2(1430)$, $K^*(1680)$, $K^*(1410)$ and $\rho(1450)$--, and replacing the running width of the 
BW by a fixed value. As total systematic uncertainty associated with the
Dalitz model, given in Table~\ref{tab:cp_coord}, 
we consider the sum square of contributions from each alternative model, where each contribution is 
evaluated from the difference between the \CP fit parameters using the alternative and the reference models. The dominant 
contributions to the overall Dalitz model uncertainty arise from the 
$\pi\pi$ and $K\pi$ S-waves, and the fixed BW width.

Experimental systematic uncertainties arise from several sources and can be found in
Table~\ref{tab:cartesian-syst}. All of them are small compared with the statistical precision, and their sum
is similar to the Dalitz model uncertainty. Other possible sources of experimental systematic uncertainty are found to be negligible.

\begin{table*}
\caption{\label{tab:cartesian-syst} Summary of the main contributions to the experimental systematic error on the \CP parameters. 
}
\begin{ruledtabular}
\begin{tabular}{lcccccccccccc}
\\[-0.15in]
 Source                               &$x_-$  & $y_-$ & $x_+$ & $y_+$  & $x^*_-$ & $y^*_-$ & $x^*_+$ & $y^*_+$ & \xsm & \ysm & \xsp & \ysp \\   [0.01in] \hline
 \mes, \DeltaE, \fis shapes           & 0.002 & 0.004 & 0.003 & 0.004  & 0.011 & 0.012 & 0.008 & 0.008         & 0.08 & 0.12 & 0.10 & 0.12  \\ 
 Real \Dz\ fractions                  & 0.002 & 0.000 & 0.000 & 0.000  & 0.002 & 0.003 & 0.002 & 0.016         & 0.03 & 0.03 & 0.03 & 0.04 \\ 
 Charge-\Dz flavor correlation        & 0.008 & 0.002 & 0.002 & 0.002  & 0.005 & 0.005 & 0.001 & 0.022         & 0.03 & 0.04 & 0.03 & 0.05 \\ 
 Efficiency in the Dalitz plot        & 0.014 & 0.000 & 0.013 & 0.001  & 0.001 & 0.002 & 0.000 & 0.001         & 0.06 & 0.04 & 0.07 & 0.09 \\ 
 Background Dalitz shape              & 0.006 & 0.003 & 0.001 & 0.004  & 0.012 & 0.015 & 0.009 & 0.009         & 0.04 & 0.09 & 0.04 & 0.09 \\
 Dalitz amplitudes and phases         & 0.004 & 0.004 & 0.004 & 0.004  & 0.008 & 0.008 & 0.008 & 0.008         & 0.01 & 0.01 & 0.01 & 0.01 \\
 $B^-\to D^{*0}K^-$ cross-feed        & 0.000 & 0.000 & 0.000 & 0.000  & 0.004 & 0.001 & 0.004 & 0.004         & --   & --   & --   & --   \\
 \CP violation in $D\pi$ and \BB bkg  & 0.000 & 0.000 & 0.000 & 0.000  & 0.005 & 0.002 & 0.002 & 0.005         & 0.00 & 0.00 & 0.00 & 0.00 \\ \hline
 Total experimental                   & 0.018 & 0.007 & 0.014 & 0.007  & 0.020 & 0.022 & 0.015 & 0.032         & 0.11 & 0.16 & 0.13 & 0.18 \\  

\end{tabular}
\end{ruledtabular}
\end{table*}

\section{Interpretation and Conclusions}

\begin{figure}[htb]
\begin{tabular}{cc}
\includegraphics[width=0.235\textwidth]{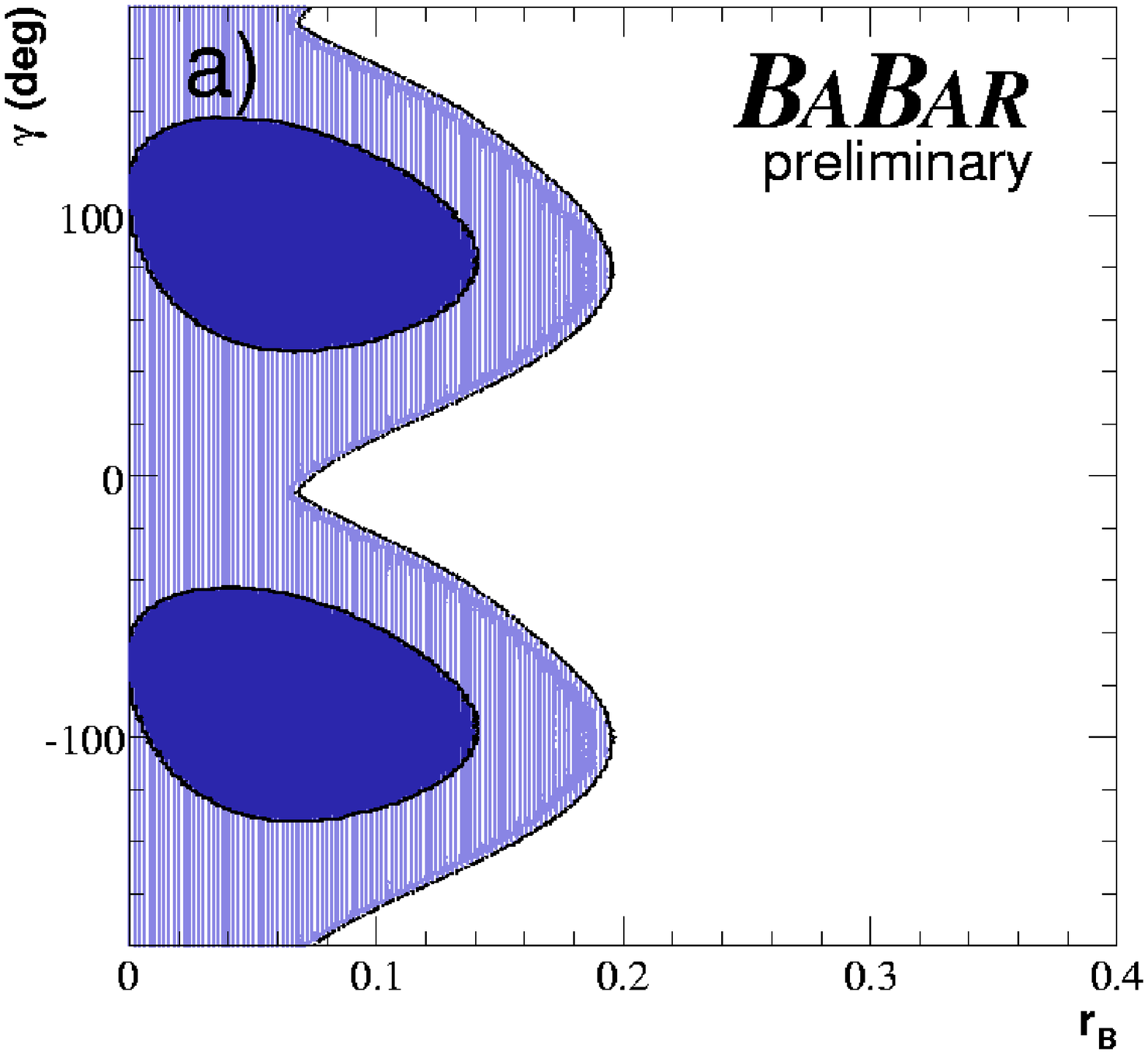}&
\includegraphics[width=0.235\textwidth]{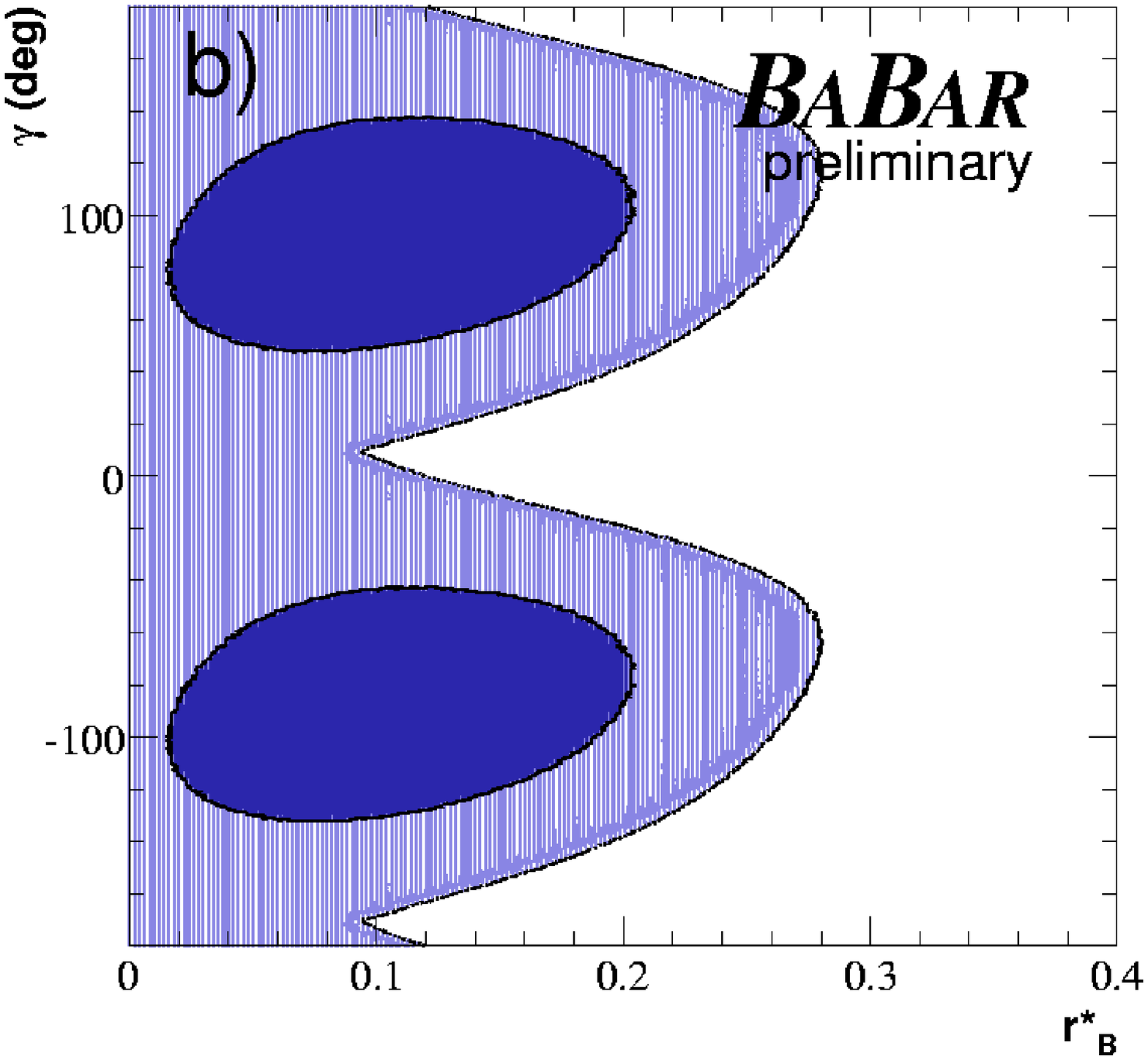}\\
\end{tabular}
\caption{\label{fig:gamma_rb_proj} Projections onto the (a) $(r_B,\gamma)$ and (b) $(r^*_B,\gamma)$
planes of the 3.7\% (dark) and 45.1\% (light) five-dimensional confidence level regions, corresponding
to one- and two-standard deviation intervals, respectively, including statistical and systematic uncertainties.}
\end{figure}

A frequentist (Neyman) 
procedure~\cite{ref:pdg2004,ref:babar_dalitzpub} 
has been adopted to interpret
the measurement of the \CP\ parameters $(x^{(*)}_\mp,y^{(*)}_\mp)$ 
reported in Table~\ref{tab:cp_coord} in terms of confidence regions on 
$(\gamma,r_B,\delta_B,r^*_B,\delta^*_B)$.
No attempt has been done yet to include in the combination the \CP parameters for $B^\mp\rightarrow D K^{*\mp}$ decays,
$(x_{s\mp},y_{s\mp})$.
Figure~\ref{fig:gamma_rb_proj} shows the two-dimensional
projections onto the $(r_B,\gamma)$ and $(r^*_B,\gamma)$ planes of the one- and two-standard deviation regions, 
including statistical and systematic uncertainties. The figure reveals the two-fold ambiguity of this method,
$(\gamma,\delta^{(*)}_{B,s}) \to (\gamma + 180^\circ, \delta^{(*)}_{B,s} + 180^\circ)$, 
as expected from Eq.~(\ref{eq:ampgen}).
From the one-dimensional projections we obtain for the weak phase
$\gamma = (92 \pm 41 \pm 11 \pm 12)^\circ$,
and for the strong phase differences  
$\delta_{\B} = (118 \pm 63 \pm 19 \pm 36)^\circ$ and 
$\delta^*_{\B} = (-62 \pm 59 \pm 18 \pm 10)^\circ$. No constraints on the phases are 
achieved at two-standard deviation level and beyond. 
Similarly, for the
magnitude of the ratio of decay amplitudes $r_\B$ and $r_\B^*$ we obtain
the one- (two-) standard deviation constraints $r_\B<0.140~(r_\B<0.195)$ and $0.017<r_\B^*<0.203~(r_\B^*<0.279)$.
No constraint on $\gamma$ is obtained from $B^\mp\rightarrow D K^{*\mp}$ decays alone, for which
$\kappa r_s<50 (0.75)$ at one- (two-) standard deviation level.
All these results are obtained considering the statistical correlations discussed in Sec.~\ref{sec:cpfit},
while the experimental and Dalitz model systematic uncertainties are taken uncorrelated. 
We have verified that accounting for experimental systematic correlations within
a given $B$ decay channel, $(x_\mp,y_\mp)$, $(x^*_\mp,y^*_\mp)$, or $(x_{s\mp},y_{s\mp})$, or assuming 
the experimental and Dalitz model systematic uncertainties between
$(x_\mp,y_\mp)$ and $(x^*_\mp,y^*_\mp)$ fully correlated, has a negligible effect on the results.

In conclusion, \babar\ has reached a good precision in the measurement of the \CP parameters $(x^{(*)}_\mp,y^{(*)}_\mp)$ but
the improvement of the statistical and systematic uncertainties on $\gamma$ also depends on the value
of the $r_B^{(*)}$ parameter (the former scales as $1/r^{(*)}_B$).
Our last experimental results for $r^{(*)}_B$ are somewhat smaller than in our 
previously published analysis~\cite{ref:babar_dalitzpub}. 
Therefore, an improved precision in the determination of $r^{(*)}_B$ is fundamental to better constraint $\gamma$. 
In this respect, the analysis of more data being recorded by the detector,
the addition of different $B$ (e.g. $B^\mp\rightarrow D K^{*\mp}$) and $D$ (e.g. $\Dz \to \piz \pim \pip, \KS K^- K^+$) decay channels, 
and the combination with other methods~\cite{ref:gronau,ref:soni} will be helpful. Assuming $r_B = 0.1$ it will
be possible to measure $\gamma$ with $\sim 10^\circ$ error with a 1~\invab data sample, which is within the reach of the \babar\ experiment.

\begin{acknowledgments}
I wish to thank the Spanish MEC under grant FPA2005-05142 for the support for this work,
and M.~Rama and N.~Neri for reading the manuscript and useful comments.
\end{acknowledgments}


\begin{thebibliography}{99}
\bibitem{ref:CKM} N.~Cabibbo, Phys. Rev. Lett. {\bf 10}, 531 (1963); 
                  M.~Kobayashi and T.~Maskawa, Prog. Theor. Phys. {\bf 49}, 652 (1973).   

\bibitem{ref:wolfenstein} L.~Wolfenstein, \jprl{51}, 1945 (1983).
                   
\bibitem{ref:chargeconj} Reference to the charge-conjugate state is implied here and throughout the text unless otherwise specified.

\bibitem{ref:gronau} M.~Gronau and D.~London, \plb{253}, 483 (1991); 
                     M.~Gronau and D.~Wyler, \plb{265}, 172 (1991);

\bibitem{ref:soni}  D.~Atwood, I.~Dunietz and A.~Soni, \jprl{78}, 3257 (1997).

\bibitem{ref:ggsz_ads} A.~Giri, Y.~Grossman, A.~Soffer and J.~Zupan, \jprd{68}, 054018 (2003).

\bibitem{ref:belle_dal04} Belle Collaboration, A.~Poluetkov {\em et al.}, \jprd{70}, 072003 (2004).


\bibitem{ref:dmixing} Y.~Grossman, A.~Soffer, J.~Zupan, \prd{72}, 031501 (2005).

\bibitem{ref:dcpv} CLEO Collaboration, D.~M.~Asner {\it et al.}, \jprd{70}, 091101 (2004).

\bibitem{ref:babar_dalitzpub} \babar\ Collaboration, B.~Aubert {\em et al.}, \jprl{95}, 121802 (2005). 

\bibitem{ref:bondar_gershon} A.~Bondar and T.~Gershon, \jprd{70}, 091503 (2004).

\bibitem{ref:gronau2002} M.~Gronau, Phys. Lett. {\bf B557}, 198 (2003).

\bibitem{ref:babar} \babar\ Collaboration, B.\ Aubert {\em et al.}, Nucl.\ Instrum.\ Methods {\bf A479}, 1-116 (2002).

\bibitem{ref:babar_dalitzeps05} \babar\ Collaboration, B.~Aubert {\em et al.}, hep-ex/0507101. 

\bibitem{ref:babar_dalitzeps06} \babar\ Collaboration, B.~Aubert {\em et al.}, hep-ex/0607104. 

\bibitem{ref:blatt-weisskopf} J.~M.~Blatt, V.~F.~Weisskopf, ``Theoretical Nuclear Physics'', John Wiley \& Sons, New York (1952). 

\bibitem{ref:gounarissakurai} G.J.~Gounaris and J.J.~Sakurai, \jprl{21}, 244 (1968).

\bibitem{ref:cleo} CLEO Collaboration, S.~Kopp {\it et al.}, \jprd{63}, 092001 (2001).

\bibitem{ref:pdg2004} Particle Data Group, S.~Eidelman {\em et al.}, \plb{592}, 1 (2004).

\bibitem{ref:e791K*} E791 Collaboration, E.~M.~Aitala {\it et. al.}, \jprl{89}, 121801 (2002).

\bibitem{ref:comment_sigma} The $\sigma$ and $\sigma'$ masses and widths are determined 
from the data. We find (in~\mevcc) $M_{\sigma}=490\pm6$, $\Gamma_{\sigma}=406\pm11$, 
$M_{\sigma'}=1024\pm4$, and $\Gamma_{\sigma'}=89\pm7$. 
Errors are statistical.

\bibitem{ref:Kmatrix} 
                      I.~J.~R.~Aitchison, \npa{189}, 417 (1972).

\bibitem{ref:multimeson} Multi-meson channel refers to a final state with four pions. 

\bibitem{ref:AS}  V.V.~Anisovich and A.V.~Sarantsev, \epj{A16}, 229 (2003).

\bibitem{ref:lassparam1} LASS Collaboration, D.~Aston {\em et al.}, \npb{296}, 493 (1988).

\bibitem{ref:zemach} V.~Filippini, A.~Fontana and A.~Rotondi, \jprd{51}, 2247 (1995).


\end{thebibliography}
\end{document}